\documentclass[10pt,twocolumn,twoside,letterpaper]{IEEEtran}
\IEEEoverridecommandlockouts

\usepackage{cite}
\usepackage{amsmath,amssymb,amsfonts}
\usepackage{algorithmic}
\usepackage{graphicx}
\usepackage{textcomp}
\usepackage{xcolor}

\usepackage{geometry}
\makeatletter
\def\ps@IEEEtitlepagestyle{
	\def\@oddfoot{\mycopyrightnotice}
	\def\@evenfoot{}
}
\def\mycopyrightnotice{
	{\footnotesize
		\begin{minipage}{\textwidth}
			\centering
			978-1-7281-4164-0/19/\$31.00 \copyright2019 IEEE
		\end{minipage}
	}
}

\begin{document}
\title{Processing of Compton events in the PETALO readout system
\thanks{This work was supported by the European Research Council (ERC) under grant ID 757829, the Generalitat Valenciana under grants PROMETEO/2016/120 and SEJI/2017/011, and the Spanish Ministry of Economy and Competitiveness under project FPA2016-78595-C3-3-R.}%
}

\author{PETALO Collaboration}%
\author{J. Renner, J.M. Benlloch-Rodr\'{i}guez, J.V. Carri\'{o}n, R. Gadea,\\
	V. Herrero-Bosch, M. Kekic, C. Romo-Luque, P. Ferrario, and J. J. G\'omez-Cadenas%
	\thanks{J. Renner (corresponding author, email: jedward.renner@usc.es) is with the Instituto Gallego de F\'isica de Altas Energ\'ias (IGFAE), Univ.\ de Santiago de Compostela, Santiago de Compostela, Spain, and the Instituto de F\'isica Corpuscular (IFIC), CSIC \& Universitat de Val\`encia, Paterna, Spain.}
	\thanks{J.M. Benlloch-Rodr\'{i}guez, J.V. Carri\'{o}n, M. Kekic, and C. Romo-Luque are with the Instituto de F\'isica Corpuscular (IFIC), CSIC \& Universitat de Val\`encia, Paterna, Spain.}
	\thanks{R. Gadea and V. Herrero-Bosch are with the Instituto de Instrumentaci\'on para Imagen Molecular (I3M), CSIC \& Universitat Polit\`ecnica de Val\`encia, Valencia, Spain.}%
	\thanks{P. Ferrario and J. J. G\'omez-Cadenas are with the Donostia International Physics Center (DIPC), Donostia-San Sebastian, Spain, and the Basque Foundation for Science (IKERBASQUE), Bilbao, Spain.}%
}

\maketitle

\begin{abstract}
PETALO (Positron Emission TOF Apparatus based on Liquid xenOn) exploits the unique characteristics of liquid xenon as a scintillator for use in a PET detector. Here initial simulation studies are detailed which highlight the potential of such a detector and outline the steps taken in the reconstruction of 511 keV gamma rays and in the full PET image reconstruction. In particular, a neural network-based approach is conceived in order to tag gamma rays that are poorly reconstructed due to Compton scattering. It is found that though a significant fraction of events undergo Compton scattering, not all of these events will necessarily be poorly reconstructed. Further study is necessary to determine whether or not an online implementation of the neural network-based event tagging will provide sufficient recovery of the image quality.
\end{abstract}

\begin{IEEEkeywords}
PET, liquid xenon, tomography
\end{IEEEkeywords}

\section{Introduction}
\IEEEPARstart{L}{iquid} xenon presents several advantages as a detection medium in PET.  Its fast time response and high scintillation yield allow for the construction of large area detectors with customizable shape which are required for full body PET scanners. The PETALO (Positron Emission TOF Apparatus based on Liquid xenOn) continuous detector concept is based on a cylindrical structure with its outer wall covered by a densely packed array of VUV sensitive silicon photomultipliers (SiPMs) and a layer of liquid xenon acting as scintillator medium. 

In order to benefit fully from PETALO's capabilities a readout architecture was designed to handle the large amount of data generated by the detector without degrading timing characteristics. Image compression techniques have been applied to relieve data link speed requirements making readout more affordable in terms of cost and complexity. However the true potential of the continuous detector has not yet been fully exploited. A deeper processing of the light distribution acquired by the detector could enhance its performance. Monte Carlo analysis of the gamma ray detection process shows a remarkable amount of events that, although affected by Compton scattering, introduce a negligible deviation in final position estimation. In this paper a novel processing technique based on a hardware-implemented neural network inside the readout is proposed. This new element would provide a quality tagging mechanism which measures the potential position reconstruction error due to Compton scattering during event detection. This information would be available in the first stage of the data acquisition system, allowing for early rejection of the worst events thus further reducing data traffic. Also reconstructed image quality may be improved by adding a confidence level parameter to the accepted events.

\begin{figure}[htbp]
	\centerline{\includegraphics[width= 0.35\textwidth]{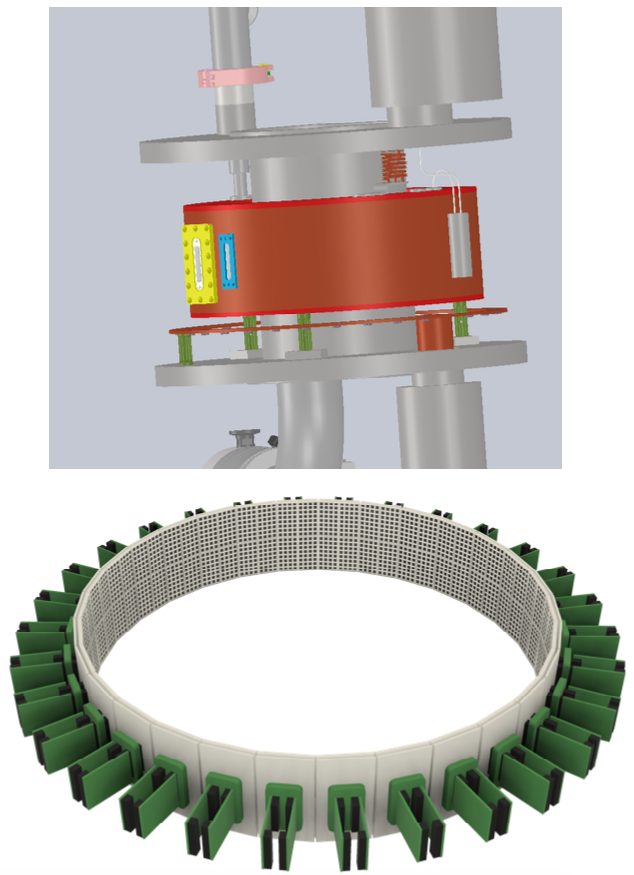}}
	\caption{PETALO cryostat (above) and outer surface instrumented with SiPMs (bottom).}
	\label{fig.geom}
\end{figure}

\section{The PETALO Detector}
The PETALO detector geometry \cite{petalo_IEEE2018} consists of a ring-shaped volume filled with liquid xenon cooled by a cryostat (see Figure \ref{fig.geom}). One or both inward-facing surfaces may be instrumented with SiPMs, which capture most of the scintillation light with fewer distortions due to border effects than other continuous scintillator based detectors \cite{Lerche_2008}. For optimal performance, PETALO uses densely packed arrays of VUV sensitive SiPMs that allow for maintaining LXe timing properties and give better resolution specifications compared to previous LXe-based detectors \cite{Nishikido_2004}.

In order to take advantage of PETALO's potentially outstanding characteristics, a readout architecture has been designed with the following specifications in mind:
\begin{enumerate}
	\item[\textbullet] The readout must be fully expandable to any detector size with minimum cost in complexity.
	\item[\textbullet] The architecture of the readout must be compatible with the continuous design of the detector.
	\item[\textbullet] The amount of information generated per detected event is much higher than that of a segmented detector, and data acquisition electronics must be able to handle this.
	\item[\textbullet] The time of flight (TOF) resolution must be the same for any detector size, and degradation due the behavior of the electronics should remain negligible compared to the intrinsic detector resolution.
\end{enumerate}

\begin{figure}[htbp]
	\centerline{\includegraphics[width= 0.45\textwidth]{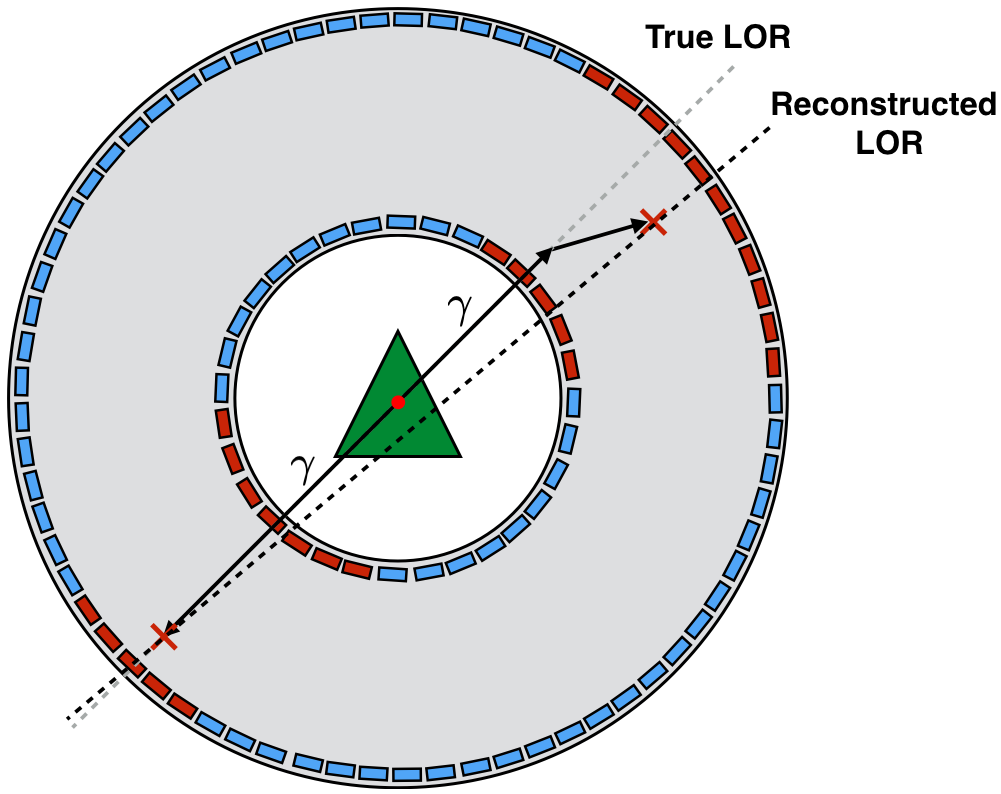}}
	\caption{Compton scattering within the active volume.}
	\label{fig.Compton}
\end{figure}

One potential issue in a LXe-based PET is Compton scattering. An incident 511 keV gamma ray may undergo Compton scattering,
\begin{enumerate}
	\item[a)] \emph{outside the active region.} In this case the observed gamma ray will have less than 511 keV total energy.  Such events can be eliminated by an energy cut, and therefore the efficiency with which this can be done is dependent on the energy resolution of the detector.
	\item[b)] \emph{inside the active region.} For these gamma rays the entire energy will be observed, but the pattern of scintillation observed on the SiPMs will not be that of a single pointlike interaction. This could affect the reconstructed location of the interaction and therefore give an improperly reconstructed line of response (LOR), as detailed in Fig. \ref{fig.Compton}.
\end{enumerate}

\begin{figure}[htbp]
	\centerline{\includegraphics[width= 0.45\textwidth]{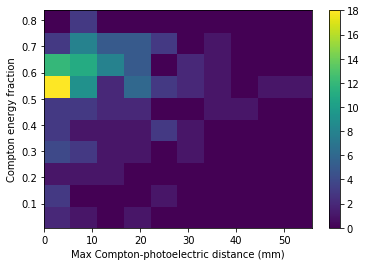}}
	\caption{Fraction of total energy deposited in Compton scatter vs. maximum distance traveled between two Compton scatters in LXe.}
	\label{fig.Compton_fraction}
\end{figure}

Compton scattering is common ($\sim$ 80\% of 511 keV gamma ray interactions) in LXe, and can vary in severity. For example, a gamma ray that performs a single Compton scatter and then proceeds to deposit the remainder of its energy in the active volume may travel a short distance (negligible compared to the resolution of PETALO), or it may travel a much longer distance (several cm). It may also deposit most of its energy in the Compton scatter, or very little. As shown in Fig. \ref{fig.Compton_fraction}, it is most likely that a 511 keV gamma ray leaves of order half of its energy in the initial interaction and travels less than 1 cm before interacting again. However, some leave very little energy in the initial scatter, and some travel a significant distance before interacting again. A simulation-based study was conducted to investigate how this effects the operation of PETALO.

\begin{figure*}[htbp]
	\centerline{\includegraphics[width= 0.8\textwidth]{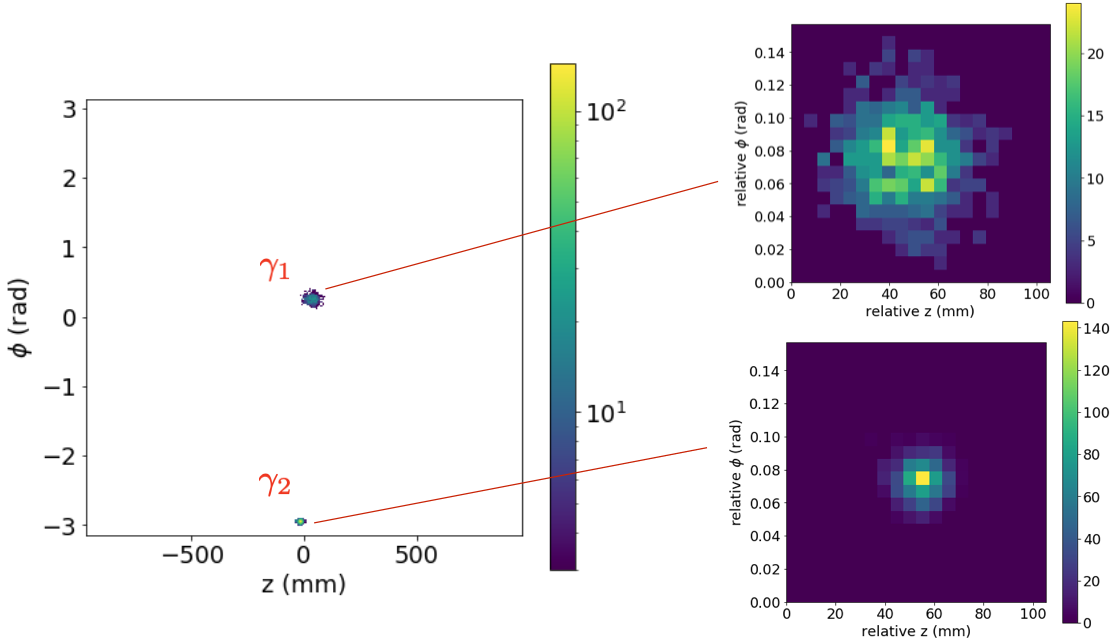}}
	\caption{SiPM responses for an event in which two 511 keV gamma rays were detected in coincidence.}
	\label{fig.gamma_reco}
\end{figure*}

\section{Simulation}\label{sec.simulation}
To drive the initial studies presented here, a Geant4-based \cite{Geant} simulation was performed for a full-body sized PETALO detector geometry. This was a ring of about 1.94 m in extent (along the $z$-axis) and inner and outer radii of approximately 38 and 41 cm filled with LXe. The inside of the outer ring was instrumented with SiPMs placed in 278 rings of 368 sensors with a 7 mm pitch. In the simulation, back-to-back 511 keV gamma rays were emitted from individual points chosen randomly according to a 3D phantom. The scintillation produced when the gamma rays interacted in the xenon was recorded by the SiPMs, and the resulting light patterns were saved along with the true energy deposited and timing information. Events in which the total measured energy in the SiPMs corresponding to each gamma ray deposition was found to be between within a suitable range (1050 and 1300 photoelectrons) were considered coincidences. An example of the SiPM signals corresponding to a coincidence is shown in Fig. \ref{fig.gamma_reco}. For each coincidence, the $(z,\phi)$ positions of the two gamma rays were reconstructed as 

\begin{equation}\label{eqn.zphi}
	z = \frac{1}{Q_{z}}\sum_{q_{i} > q_{t,z}}z_{i}q_{i},\,\,\,\, \phi = \frac{1}{Q_{\phi}}\sum_{q_{i} > q_{t,\phi}}\phi_{i}q_{i},
\end{equation}

\noindent where $q_{i}$, $z_{i}$, and $\phi_{i}$ are the charge, z-coordinate, and $\phi$-coordinate of SiPM $i$ and $q_{t,z}$ and $q_{t,\phi}$ are charge thresholds for including the contribution of SiPM $i$, both set to 4 photoelectrons in this study. $Q_{z}$ and $Q_{\phi}$ are the total SiPM charges contributing to the calculation of the respective coordinate, 

\begin{equation}
Q_{z} = \sum_{q_{i} > q_{t,z}}q_{i},\,\,\,\, Q_{\phi} = \sum_{q_{i} > q_{t,\phi}}q_{i}.
\end{equation}

\begin{figure}[htbp]
	\centering
	\includegraphics[width= 0.4\textwidth]{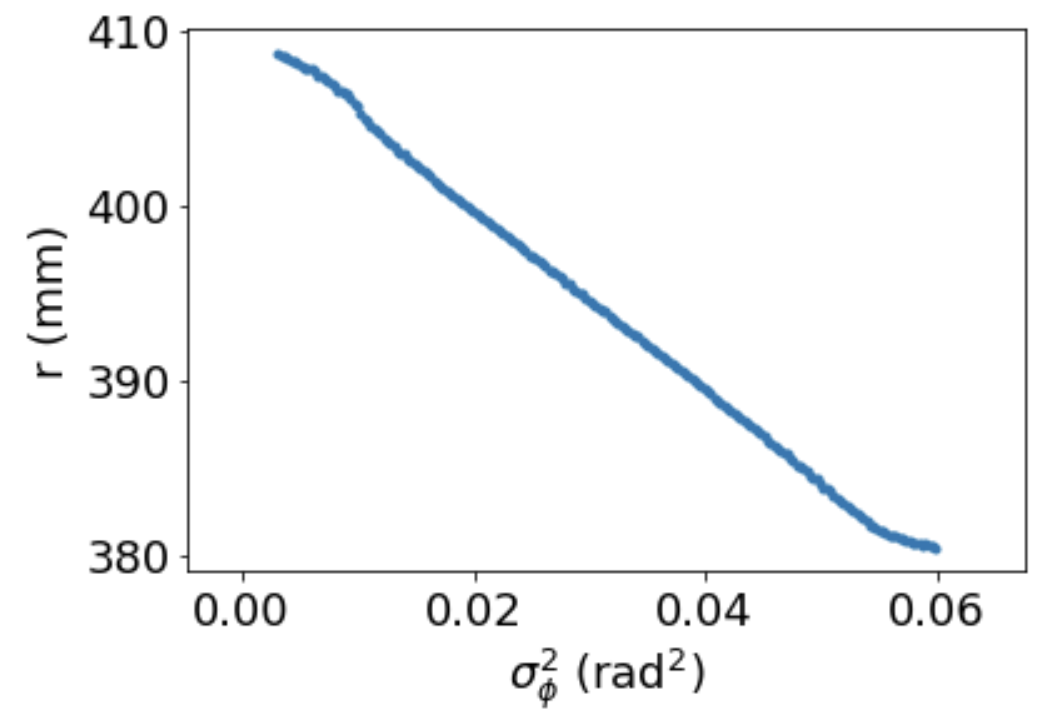}
	\caption{Profile of the radial coordinate vs. the variance of the coordinate $\phi$. To reconstruct the radial coordinate, $r = f_{r}(\sigma_{\phi}^2)$ is read off of this curve given $\sigma_{\phi}^2$ determined from the sensor information.}
	\label{fig.fr}
\end{figure}

\noindent The radial coordinate $r$ was computed as

\begin{equation}\label{eqn.r}
	r = f_{r}(\sigma^{2}_{\phi}),\,\,\,\, \mathrm{for}\,\,\,\, \sigma^{2}_{\phi} = \frac{1}{Q_{\phi}}\sum_{q_{i} > q_{t,\phi}}(\phi_{i} - \phi)^{2}q_{i},
\end{equation}

\noindent with $\phi$ as defined in equation \ref{eqn.zphi}. The function $f_{r}(\sigma^{2})$ which yields the radial coordinate as a function of variance in $\phi$ is shown in Fig. \ref{fig.fr}.

\begin{figure*}[htbp]
	\hspace{0.7cm}Simulated Phantom\hspace{3.5cm}True + TOF\hspace{4.2cm}True, no TOF\\
	\centering
	\includegraphics[width= 0.32\textwidth]{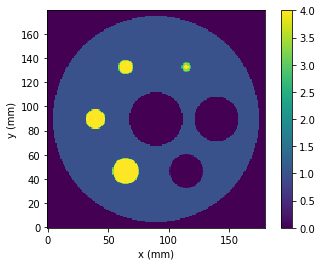}
	\includegraphics[width= 0.32\textwidth]{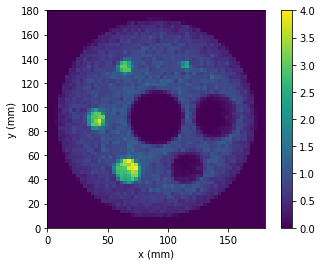}
	\includegraphics[width= 0.32\textwidth]{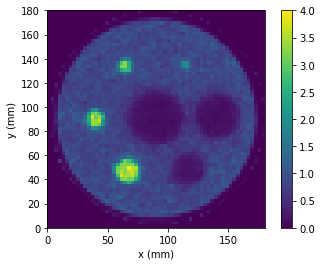}\\
	\raggedright\hspace{5.2cm}Reco + TOF\hspace{4.2cm}Reco, no TOF\\
	\centering
	\includegraphics[width= 0.32\textwidth]{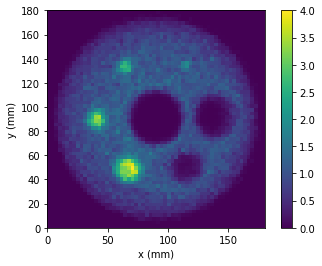}
	\includegraphics[width= 0.32\textwidth]{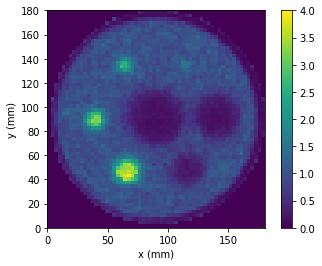}\\
	\caption{Preliminary reconstruction in a full-body PETALO detector. A 5 mm slice of the simulated phantom is shown (top left). Reconstruction based on $\sim$850 000 simulated LORs is shown using the true gamma ray interaction points with TOF (top middle) and without TOF (top right), and the randomly generated reconstructed gamma ray locations with TOF (bottom left) and without TOF (bottom right).}
	\label{fig.image_reco}
\end{figure*}

The simulated phantom was similar to the NEMA \cite{NEMA} IQ phantom, consisting of four ``hot'' spheres of radii 5, 6.5, 8.5 and 11 mm and two ``cold'' spheres of radii 14 and 18.5 mm. All spheres were placed with centers at z = 0 and at a radius of 114.4 mm from the origin. A ``cold'' cylinder of length 40 mm and radius 22 mm was placed, also centered at the origin. These entities were enclosed in a ``background'' sphere of radius 85 mm, centered at the origin. The activities of the different regions were assigned as follows: ``hot'' = 4, ``cold'' = 0, and ``background'' = 1.  The phantom was constructed with a resolution of 1 mm, and a 5 mm z-slice centered at z = 0 is shown in Fig. \ref{fig.image_reco}, top left.

\begin{figure}[htbp]
	\centering
	\includegraphics[width= 0.37\textwidth]{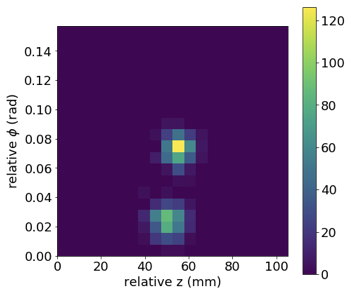}
	\caption{An example of a single-gamma (511 keV) deposition in which Compton scattering lead to two clearly separated energy depositions.}
	\label{fig.compton}
\end{figure}

\begin{figure*}[htbp]
	\includegraphics[width= 0.33\textwidth]{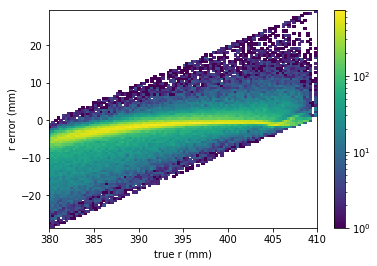}
	\includegraphics[width= 0.33\textwidth]{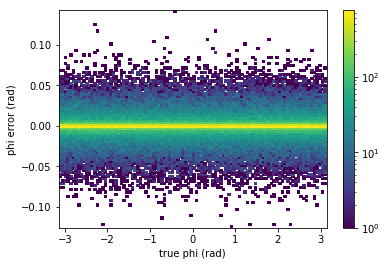}
	\includegraphics[width= 0.33\textwidth]{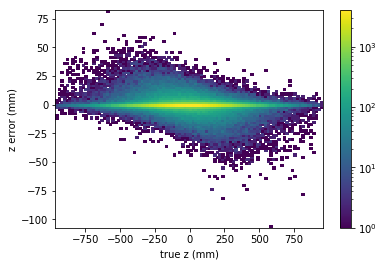}
	\includegraphics[width= 0.33\textwidth]{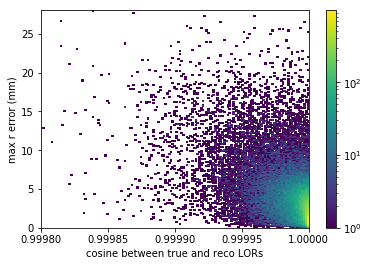}
	\includegraphics[width= 0.33\textwidth]{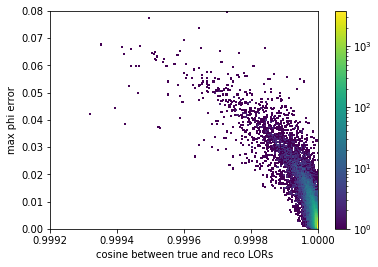}
	\includegraphics[width= 0.33\textwidth]{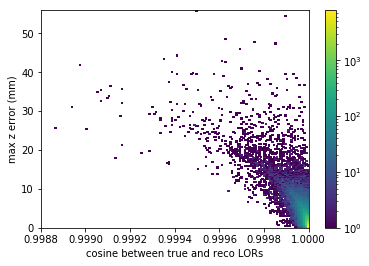}	
	\caption{Error matrices (true $-$ reconstructed) for coordinates $r$, $\phi$, and $z$ from Geant4 simulated coincidence (above). Note that only individual gammas for which the majority ($\geq$ 99.5\%) of the total observed SiPM charge could be enclosed in a 20x20 window (this cut eliminated the gamma depositions that were greatly spread out due to, for example, Compton scattering; it eliminated about 4\% of the total number of gamma depositions). Cosine of the angle between the true and reconstructed LORs vs. error (true $-$ reconstructed) in each coordinate $r$, $\phi$, and $z$ (below). Note that the error is the maximum for each coordinate of those computed for each of the two gammas involved in determining the LOR. Furthermore, for each coordinate, the errors in the other two coordinates have been restricted: for $r$, only events with $-0.02 < \phi-\mathrm{error} < 0.02$ and $5\,\mathrm{mm} < z-\mathrm{error} < 5\,\mathrm{mm}$ are shown; for $\phi$, only events with $5\,\mathrm{mm} < r-\mathrm{error} < 5\,\mathrm{mm}$ and $5\,\mathrm{mm} < z-\mathrm{error} < 5\,\mathrm{mm}$ are shown; for $z$, only events with $-0.02 < \phi-\mathrm{error} < 0.02$ and $5\,\mathrm{mm} < r-\mathrm{error} < 5\,\mathrm{mm}$ are shown.}
	\label{fig.errmat}
\end{figure*}

The errors in the reconstructed coordinates $(r,\phi,z)$ for individual gamma rays are shown in Fig. \ref{fig.errmat} (above). Note the log scale for the intensity. Poorly reconstructed events are likely to be due to Compton scattering producing a non-pointlike light distribution on the SiPMs (see Fig. \ref{fig.compton}). However, because of the range of possibilities in the amount of energy deposited in the Compton scatter and in the distance traveled before interacting again, not all events will show two clear sites of energy deposition. Some Compton scatters may even give a large error in one of the reconstructed coordinates but not appreciably affect the LOR. Figure \ref{fig.errmat} examines this by showing the cosine of the angle between the true and reconstructed LORs vs. the maximum error (out of each of the two gammas involved in determining the LOR) in each coordinate, with the errors in the other two coordinates restricted. The plot shows that independently varying the $r$-error does not necessarily affect the reconstruction of the LOR, while reconstruction errors in $\phi$ and $z$ alone are more likely to cause a misaligned LOR. This is likely due to ``soft'' Compton scatters in which very little energy is left in the initial scatter, and a second photoelectric interaction is detected after the gamma has traveled some distance in $r$. The Compton scattering angle is in this case small, and therefore the LOR is not significantly affected, despite a large error in $r$.

In section \ref{sec.NN} we discuss a neural-network based approach for identifying potentially problematic Compton scatters, and in section \ref{sec.reco} we take an initial look at how such events may affect the final reconstruction.

\begin{figure*}[htbp]
	\includegraphics[width= 0.33\textwidth]{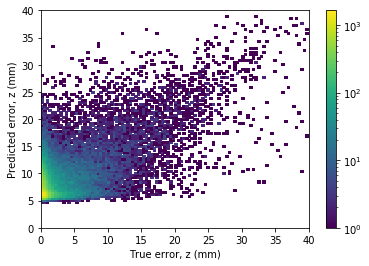}
	\includegraphics[width= 0.33\textwidth]{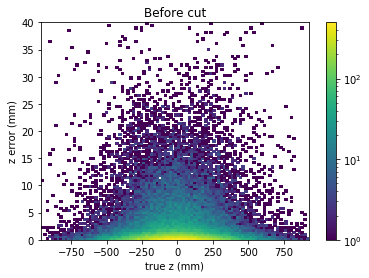}
	\includegraphics[width= 0.33\textwidth]{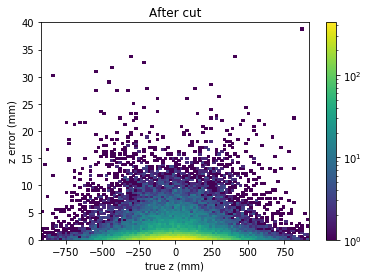}
	\caption{True magnitude in $z$-error for reconstructed individual gamma rays vs. predicted value given by the trained neural network for 48700 test events (left). True magnitude of $z$-error vs. true $z$-position for these events (middle), and the same for the events remaining after application of a cut on the predicted error of $|z\mathrm{-error}|_{\mathrm{predicted}} < 12$.}
	\label{fig.nnevaluation}
\end{figure*}

\section{Neural Network-Based Error Prediction}\label{sec.NN}
A wide variety of Compton scattering scenarios exist, and it is difficult to draw a clear line between a Compton scatter that is harmless to the reconstruction and one that is significant. Therefore we have attempted to train a neural network to predict the error in a reconstructed coordinate. With such a tool we can later decide how much error is too much and attempt to remove events that the network claims will produce such an error. From Fig. \ref{fig.errmat} (bottom), we can see that an error in $\phi$ or $z$ is more likely to indicate a poorly reconstructed LOR, so in the following example we consider the coordinate $z$.

\begin{figure}[htbp]
	\includegraphics[width= 0.5\textwidth]{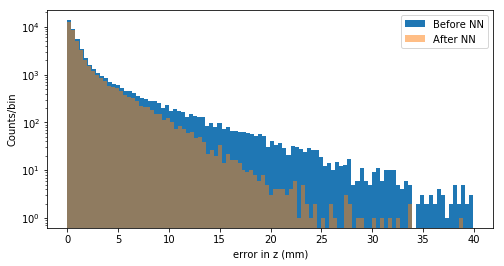}
	\caption{True magnitude in the $z$-error for reconstructed individual gamma rays for events before and after a cut on the neural-network-predicted error of $|z\mathrm{-error}|_{\mathrm{predicted}} < 12$.}
	\label{fig.nnhist}
\end{figure}

To train the network, the SiPM responses were split into two regions corresponding to the two 511 keV gamma interactions, and from each region a 20x20 window was placed around the most active SiPMs (as in Fig. \ref{fig.gamma_reco}). Gamma rays which had more than 99.5\% of their charge outside this 20x20 window (about 4\%) were discarded. A fully-connected neural network was constructed, consisting of a 400-value input layer (the flattened 20x20 array of SiPM responses), four layers of 64, 32, 16, and 8 neurons respectively, and an output layer of 1 neuron which gave the predicted value of the magnitude of the reconstructed $z$-error (normalized to the range of 0-1). The network was built and trained using Keras \cite{Keras} (Tensorflow \cite{Tensorflow} backend) with 175000 training samples, 5\% of which were used as a validation set. The predicted $z$-error magnitudes for 48700 gamma rays not included in the training/validation phase are shown vs. the magnitude of the true $z$-error in Fig. \ref{fig.nnevaluation}, along with the reconstructed $z$-error magnitude vs. true coordinate $z$ before and after a cut on the predicted error of $|z\mathrm{-error}|_{\mathrm{predicted}} < 12$ mm. The cut reduced the total number of gamma rays by about 15\%. A histogram of the $z$-error magnitudes before and after the cut is also shown in Fig. \ref{fig.nnhist}.

These results show that a neural network can be used to significantly reduce the number of poorly reconstructed gamma depositions while removing relatively fewer well-reconstructed depositions. However, further quantification using methods beyond those presented in section \ref{sec.reco} will be required to determine if such a tool would actually be useful in the overall reconstruction. Should such a tool prove beneficial, the end goal would be to develop a system in which the network is integrated into the readout electronics and performs tagging of individual gamma ray depositions on-the-fly. Initial attempts at implementing neural networks on an FPGA using the PYNQ \cite{PYNQ} development board and code adapted from the spooNN project \cite{spooNN} have shown that such a network is likely to execute fast enough ($\sim$ 10 $\mu$s) to operate online as part of the PETALO readout electronics. 

\section{Preliminary Reconstruction}\label{sec.reco}
Here we describe preliminary results on the reconstruction of a single $z$-slice of the simulated NEMA IQ phantom. The main goals of this initial work were to get a general idea of:

\begin{enumerate}
	\item[\textbullet] the extent to which Compton scattering will negatively affect the reconstruction in PETALO, and
	\item[\textbullet] to what extent the inclusion of TOF information would improve the reconstruction.
\end{enumerate}

To obtain enough coincidences for a reasonable reconstruction, a fast Monte Carlo was developed which generated random points of emission according to the same phantom as described in section \ref{sec.simulation} but used the error matrices shown in Fig. \ref{fig.errmat} (top) to cast random values for the reconstructed coordinates rather than simulating the full propagation of the gamma rays, generation and detection of scintillation, and sensor-based reconstruction. In particular, for each emission point a random emission direction was determined, and two random radii were chosen according to the distribution of the radial coordinate determined in the Geant4 simulation. From this information, the ``true'' interaction points of the two gammas could be determined, and random values for the errors on the coordinates $(r,\phi,z)$ for each gamma ray were cast using the error matrices. In this way, approximately 850\,000 coincidences were generated. Note that the angle of the gamma ray emission with the z-axis was restricted so that the magnitude of its cosine was less than 0.309, corresponding to an ``opening angle'' of $0.1\pi$. This was done because only LORs whose angle with the $z$-axis met this restriction were included in the sinograms.

The reconstruction was performed using STIR \cite{STIR} and the Python-STIR interface SIRF \cite{SIRF}. In particular, pre-release versions of STIR and SIRF were used which allowed for the inclusion of TOF in the reconstruction as implemented in \cite{Efthimiou_2019}. Sinograms were produced by computing the intersection of each LOR with 64 $\phi$-planes spanning axial angles from 0 to $\pi$. The planes were centered about $z=0$ and had a thickness of 5 mm in $z$. The distance, perpendicular to the $z$-axis, between the point of intersection and the center of the plane was recorded in 180 bins. When TOF was employed, sinograms were computed for 13 different TOF bins for TOFs in the range of $\pm$ 2 ns. The ``true'' TOF was used, i.e. without simulation of the procedure of computing the TOF from the sensor responses, and was calculated as the difference in distances traveled by the two gamma rays divided by the speed of light. A resolution of 200 ps was assumed in the reconstruction algorithm.

\begin{figure}[htbp]
	\centering
	True + TOF\\
	\includegraphics[width= 0.42\textwidth]{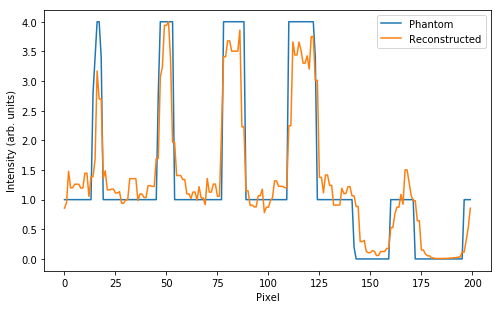}\\
	True, no TOF\\
	\includegraphics[width= 0.42\textwidth]{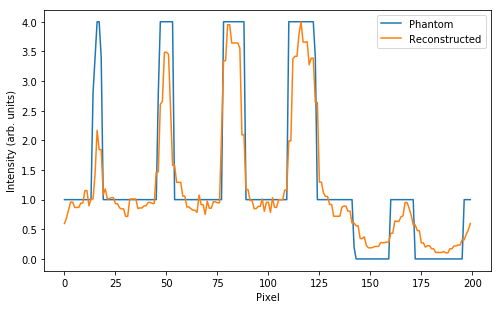}\\
	Reco +  TOF\\
	\includegraphics[width= 0.42\textwidth]{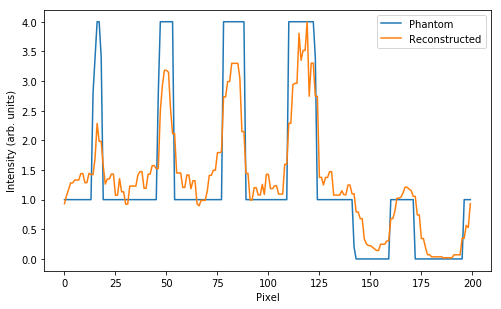}\\
	Reco, no TOF\\
	\includegraphics[width= 0.42\textwidth]{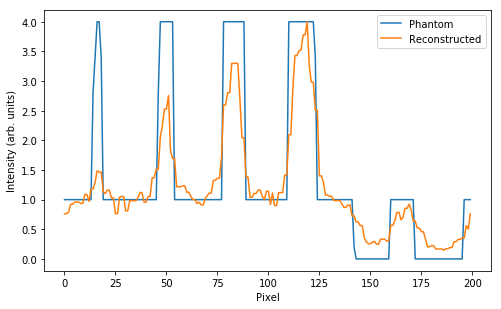}
	\caption{Radial profiles of the reconstructed images in Fig. \ref{fig.image_reco}, in which the values of the pixels are shown along a circle passing through the centers of the 5 spheres placed at $r = 50$ cm in the phantom. In each case the profile was normalized to a maximum intensity of 4 and is shown with the true profile for the simulated phantom.}
	\label{fig.graphs}
\end{figure}

Figure \ref{fig.image_reco} shows reconstructed images for 20 iterations of Ordered Subset Expectation Maximization (OSEM) with 4 subsets. The images were reconstructed with a size of 60x60, corresponding to 3 mm bin width in each dimension. The simulated distribution is shown along with the result assuming no errors in the reconstruction of the individual gamma ray interaction points with TOF information (True + TOF) and without TOF (True, no TOF), and assuming the error matrices shown in Fig. \ref{fig.errmat} (top) with TOF (Reco + TOF) and without TOF (Reco, no TOF). While it is clear that using the true gamma interaction positions produces the sharpest image, it is not dramatically different qualitatively than that produced with the reconstructed interaction points and TOF. This can also be seen by examining the radial profiles shown in Fig. \ref{fig.graphs}. The inclusion of TOF alone also seems to produce a clearer image. However, these results will need to be quantified more rigorously to measure more precisely the impact of TOF and Compton scattering.

\section{Conclusions}\label{sec.conclusions}
Initial simulation studies show that Compton scattering is not likely to present an unsurmountable problem in PETALO reconstruction, and that TOF is likely to significantly improve reconstruction capabilities. A proof-of-concept of a neural network-based solution for eliminating potential poorly reconstructed events has also been presented. Future studies include further quantification of these results and the development of an initial prototype, currently under construction at the time of this writing. 

\bibliographystyle{bib/IEEEtran}
\bibliography{bib/petalo}

\end{document}